\documentclass[%
reprint,
superscriptaddress,
bibnotes,
amsmath,amssymb,
aps,
prl,
longbibliography,
]{revtex4-2}

\usepackage{amssymb}
\usepackage{graphicx}
\usepackage{dcolumn}
\usepackage{bm}
\usepackage{chemformula}
\usepackage{hyperref}

\begin{document}


\title{On the multi-\textbf{q} characteristics of magnetic ground states of honeycomb cobalt oxides}

\author{Yuchen~Gu}
\affiliation{International Center for Quantum Materials, School of Physics, Peking University, Beijing 100871, China}
\author{Xianghong~Jin}
\affiliation{International Center for Quantum Materials, School of Physics, Peking University, Beijing 100871, China}
\author{Yuan~Li}
\email{yuan.li@iphy.ac.cn}
\affiliation{International Center for Quantum Materials, School of Physics, Peking University, Beijing 100871, China}
\affiliation{Beijing National Laboratory for Condensed Matter Physics, Institute of Physics, Chinese Academy of Sciences, Beijing 100190, China}
\date{\today}
\begin{abstract}
The Kitaev honeycomb model has received significant attention for its exactly solvable quantum spin liquid ground states and fractionalized excitations. For realizing the model, layered cobalt oxides have been considered a promising platform. Yet, in contrast to the conventional wisdom about single-\textbf{q} zigzag magnetic order inferred from previous studies of the \ch{Na_2IrO_3} and \ch{$\alpha$-RuCl_3} candidate materials, recent experiments on two of the representative honeycomb cobalt oxides, hexagonal \ch{Na_2Co_2TeO_6} and monoclinic \ch{Na_3Co_2SbO_6}, have uncovered evidence for more complex multi-\textbf{q} variants of the zigzag order. This review surveys on experimental strategies to distinguish between single- and multi-\textbf{q} orders, along with the crystallographic symmetries of the cobalt oxides in comparison to the previously studied systems. General formation mechanism of multi-\textbf{q} order is also briefly discussed. The goal is to provide some rationales for examining the relevance of multi-\textbf{q} order in the honeycomb cobalt oxides, along with its implications on the microscopic model of these intriguing quantum magnets.

\end{abstract}

\maketitle

\section{1. INTRODUCTION}
Quantum spin liquid (QSL) is a novel phase of matter in an interacting quantum many-body system \cite{BroholmScience2020, BalentsNature2010}. Originating from the idea of the resonating valence bond concept \cite{Anderson1973MRB}, QSL has garnered widespread interest partly due to the suggestion of P. W. Anderson that resonating valence bonds may be the essence of superconducting pairing in the cuprates \cite{Anderson1987Science}. A QSL phase does not exhibit spontaneous symmetry breaking even at the lowest temperatures, and features long-range entanglement and fractional quantum excitations \cite{Wen2019NPJ}. Low dimensionality, magnetic frustration, and strong quantum fluctuations are generally believed to facilitate the formation of QSLs \cite{ZhouRMP2017}. These considerations are fused into the Kitaev honeycomb model \cite{KitaevAP2006}.
The model, which is defined on a two-dimensional honeycomb lattice with spin-1/2 degrees of freedom and frustrated bond-dependent Ising interactions, has exactly solvable QSL ground states and fractionalized excitations.

A strategy for materializing the Kitaev model was first proposed by G. Jackeli and G. Khaliullin in spin-orbit coupled Mott insulators based on selected transition-metal ions \cite{JackeliPRL2009}. Over the past decade and a half \cite{TakagiNRP2019, MotomeJPCM2020, TrebstPRRSP2022}, extensive research efforts have gone into 5$d$-electron iridium-based \cite{ChaloupkaPRL2010}, 4$d$-electron ruthenium-based \cite{PlumbPRB2014}, and most recently 3$d$-electron cobalt-based compounds \cite{LiuPRB2018, SanoPRB2018, LiuPRL2020, Kim_2021, LiuIJMPB2021}. However, the candidate compounds almost always develop long-range magnetic order at low temperatures when the role of disorder is minimized \cite{TakagiNRP2019,LefrancoisPRB2016,BeraPRB2017,YanPRM2019}, so the presence of non-Kitaev interactions is evident \cite{ChaloupkaPRL2010, ChaloupkaPRL2013, KimchiPRB2011, SizyukPRB2014,RauPRL2014,ChaloupkaPRB2016,WinterPRB2016,RousochatzakisPRX2015}. The optimistic thinking is that, when the non-Kitaev terms are sufficiently small and only acting as perturbations to the Kitaev model, it would still be possible to suppress the long-range order formation through external tuning \cite{YadavSR2016, HickeyNC2019, LiuPRL2020}. A commonly considered non-Kitaev term is the isotropic Heisenberg interaction between nearest neighbors. In this context, the 3$d$ cobalt-based systems are considered favorable, because of a cancellation mechanism to suppress such Heisenberg interactions \cite{LiuPRL2020,LiuPRB2018,SanoPRB2018}.

Deriving from the conventional wisdom attained from studying the 4$d$ and 5$d$ materials \cite{CaoPRB2016, LiuPRB2011,YePRB2012}, and following the common practice of using single-\textbf{q} ansatz in the interpretation of diffraction data \cite{LefrancoisPRB2016,BeraPRB2017,YanPRM2019}, the magnetic ground state in some of the cobaltates had once been accepted to be the single-\textbf{q} zigzag order \cite{SongvilayPRB2020,Samarakoon2021PRB,Kim_2021}. However, recent investigations on \ch{Na_2Co_2TeO_6} and \ch{Na_3Co_2SbO_6} have uncovered evidence for multi-\textbf{q} order, which can be easily mistaken into the zigzag order with random domain distribution \cite{ChenPRB2021, LeePRB2021, ChenPRL2023, YaoPRR2023, Gu2024PRB, Hu2024PRB}.

In this mini review, we present an overview of topics pertinent to the formation mechanism of multi-\textbf{q} magnetic order in solids, ranging from crystallographic symmetry considerations to microscopic interactions, as well as experimental methods that can directly or indirectly indicate the existence of such order. It is hoped that such information will provide orientation for future work to assess the relevance of multi-\textbf{q} magnetic order on a more equal footing as single-\textbf{q} order, in candidate Kitaev materials and beyond, in order to constrain theoretical models with concrete characteristics of the magnetic ground state.

\section{2. Single-q order under lattice rotational symmetry}
The formation of magnetic order leads to spontaneous symmetry breaking from the paramagnetic phase. Accurate information about the crystallographic symmetry is thus crucial for understanding the symmetry of the magnetic ground state. It is particularly important to examine the transformation of single-\textbf{q} order's propagation vector under the crystallographic symmetry operations, in order to reveal whether it may have a multi-\textbf{q} variant. A recent review \cite{Garlea2024arxiv} provides an excellent summary of the crystallographic and magnetic symmetries for the candidate Kitaev materials.

\ch{Na_2IrO_3} \cite{ChaloupkaPRL2010,SinghPRB2010} and \ch{$\alpha$-RuCl_3} \cite{CaoPRB2016} are the most extensively studied candidate Kitaev materials in the 5$d$ and 4$d$ categories, respectively. \ch{Na_2IrO_3} is established to have a monoclinic $C2/m$ structure through x-ray diffraction on single crystals with a low concentration of stacking faults \cite{ChoiPRL2012}. In the case of \ch{$\alpha$-RuCl_3}, extensive research has consistently identified its room-temperature structure as monoclinic $C2/m$ \cite{JohnsonPRB2015, CaoPRB2016}. However, recent studies have revealed a structural phase transition from $C2/m$ to rhombohedral $R\overline{3}$ upon cooling below about 150 K \cite{MuPrr2022, Zhang2024PRM, Park2024JOP, Kim2024Arxiv}. The completeness of this transition, along with thermal transport properties in the low-temperature phase, is highly sensitive to stacking faults in the crystals \cite{Zhang2024PRM}.

\begin{figure}[!bh]
\includegraphics[width=3.3in]{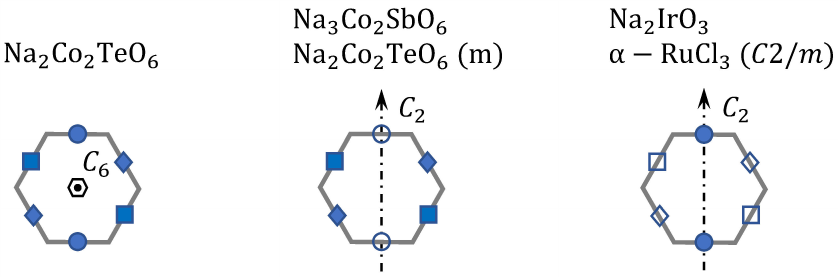}
\caption{Magnetic diffraction pattern of different compounds within the first 2D hexagonal Brillouin zone, \textit{i.e.}, without considering the $L$ components. \ch{Na_2Co_2TeO_6} features three sets of diffraction peaks which are linked by the $C_3$ rotation. These peaks are produced either together as Fourier components of a triple-$\mathbf{q}$ magnetic order, or separately by three types of single-$\mathbf{q}$ magnetic domains. Although \ch{Na_2Co_2TeO_6}~(m), \ch{Na_3Co_2SbO_6}, \ch{Na_2IrO_3} and \ch{$\alpha$-RuCl_3} all possess $C2/m$ space-group symmetry and have a $C_2$ symmetry axis along $(0, K, 0)$ at room temperature, only \ch{Na_2Co_2TeO_6}~(m) and \ch{Na_3Co_2SbO_6} have two different sets of symmetry-related magnetic diffraction peaks. Notably, the exact lattice symmetry of \ch{$\alpha$-RuCl_3} at low temperatures remains a topic of ongoing discussion.
}
\label{fig1}
\end{figure}

The propagation vectors in both \ch{Na_2IrO_3} and \ch{$\alpha$-RuCl_3} are characterized by the $M$-point of the pseudo-hexagonal two-dimensional (2D) Brillouin zone of the honeycomb lattice that is passed through by the $C_2$-symmetric \textbf{b} axis, \textbf{q} = $(0, 1, 1/2)$ \cite{YePRB2012} and $(0, 1, 1/3)$ \cite{CaoPRB2016} in the monoclinic notation, respectively (the difference is in the stacking of the ordered moments along the \textbf{c} axis). Because the 2D wave vector $(0, 1)$ is unique and invariant under the monoclinic symmetry operations (Fig.~\ref{fig1}), the single-$\mathbf{q}$ order has no multi-\textbf{q} variant, and its formation does not necessitate any spontaneous breaking of the lattice rotational symmetry. This understanding in the context of \ch{$\alpha$-RuCl_3}, however, relies on the low-temperature crystal structure being monoclinic. If the genuine crystal structure is $R\overline{3}$ \cite{MuPrr2022, Zhang2024PRM, Park2024JOP, Kim2024Arxiv}, the formation of the single-$\mathbf{q}$ order may have observable consequences related to the broken $C_3$ rotational symmetry.

\begin{table}[!h]
\begin{ruledtabular}
\renewcommand{\arraystretch}{1.3}
\begin{tabular}{lccl}
- & space group & wave vector (r.l.u.) & Refs. \\
\hline
\ch{Na_2IrO_3} & $C2/m$ & $(0, 1, 1/2)$  & \cite{YePRB2012,ChoiPRL2012} \\
\ch{$\alpha$-RuCl_3} & $C2/m$ or $R\overline{3}$ & $(0, 1, 1/3)$ ($C2/m$) & \cite{CaoPRB2016,Zhang2024PRM} \\
\ch{BaCo_2(AsO_4)_2} & $R\overline{3}$ & $(0.27, 0, -1.31)$  & \cite{Regnault1977physicaBC,HalloranPNAS2023} \\
\ch{Na_2Co_2TeO_6} & $P6_3 22$ & $(1/2, 0, 0)$  & \cite{LefrancoisPRB2016,BeraPRB2017} \\
\ch{Na_2Co_2TeO_6~(m)} & $C2/m$ & $(\pm1/2, 1/2, 0)$  & \cite{Dufault2023PRB} \\
\ch{Na_3Co_2SbO_6} & $C2/m$ & $(\pm1/2, 1/2, 0)$ & \cite{YanPRM2019} \\
\end{tabular}
\end{ruledtabular}
\caption{Space groups and magnetic propagation wave vectors of some candidate Kitaev materials. The wave vector of \ch{$\alpha$-RuCl_3} employs the monoclinic notation.}
\label{tab1}
\end{table}

The 3$d$ cobaltates can be thought of as high-spin counterparts to the 5$d$ and 4$d$ systems \cite{LiuPRB2018, SanoPRB2018, LiuIJMPB2021}. The candidate materials \ch{BaCo_2(AsO_4)_2} \cite{Regnault1977physicaBC, ZhongSA2020}, \ch{Na_2Co_2TeO_6} \cite{ViciuJSSC2007, LefrancoisPRB2016}, \ch{Na_2Co_2TeO_6}~(m) (a monoclinic polymorph) \cite{Dufault2023PRB} and \ch{Na_3Co_2SbO_6} \cite{ViciuJSSC2007,YanPRM2019} belong to the space groups $R\overline{3}$, $P6_3 22$, $C2/m$, and $C2/m$, respectively. The magnetic order in \ch{BaCo_2(AsO_4)_2} is incommensurate, with $\mathbf{q} = (0.27, 0, -1.31)$ \cite{HalloranPNAS2023, Regnault2018Helion}, distinct from the other candidate materials. Because of the complexity brought about by the incommensurability, the order's single- versus multi-\textbf{q} nature has not been closely examined in the literature. Nevertheless, we note that the system's high rhombohedral symmetry would in principle allow the order to have multi-\textbf{q} variants. The magnetic order in \ch{Na_2Co_2TeO_6} and \ch{Na_3Co_2SbO_6}, presumed here to be single-\textbf{q}, are characterized by commensurate 2D wave vectors at the $M$-point of the (pseudo)hexagonal Brillouin zone \cite{LefrancoisPRB2016, BeraPRB2017, YanPRM2019}. As shown in Fig.~\ref{fig1}, the crystal structure of hexagonal \ch{Na_2Co_2TeO_6} has $C_6$ symmetry about the \textbf{c} axis, which introduces three symmetry-related orientational domains that are energetically degenerate, all breaking the lattice rotational symmetry. The cases of monoclinic \ch{Na_3Co_2SbO_6} and \ch{Na_2Co_2TeO_6}~(m) are different. While they share the same monoclinic structural symmetry as \ch{Na_2IrO_3} and \ch{$\alpha$-RuCl_3} (at room temperature), the pertinent $M$-point is not passed through by the monoclinic $C_2$ axis \cite{YanPRM2019, LiPRX2022, Gu2024PRB} (Fig.~\ref{fig1}). As a result, the single-\textbf{q} order would break all lattice rotational symmetry. The crystallographic space group and magnetic wave vectors of different compounds are summarized in table~\ref{tab1}.

As an aside, traditionally, magnetism in candidate Kitaev materials is described based on localized magnetic moments on the honeycomb lattice, as they are formed by the atomic spin-orbit coupled $J_\mathrm{eff} = \frac{1}{2}$ pesudospins. However, such an ionic picture has been challenged in honeycomb iridium and ruthenium oxides \cite{MazinPRL2012,Foyevtsova2013PRB,StreltsovPRB2015}, based on \textit{ab initio} calculations which showed that the electronic band width due to inter-site hopping is comparable to the Hubbard repulsion $U$. It is proposed that an alternative, more accurate low-energy description of the electronic structure involves a linear combination of six $t_{2g}$ atomic orbitals on each hexagon of the honeycomb lattice, dubbed quasi-molecular orbitals (QMOs, in analogy with benzene molecules). While the precise electronic structure has remained a subject of debate \cite{Gretarsson2013PRL, MazinPRB2013, Kim2016PRL, Kim2014PRB, Igarashi2015JOP,Suzuki2019NM,LebertPRB2023}, it is argued that the magnetic order in \ch{Na_2IrO_3} can also be understood from the QMO point of view \cite{MazinPRL2012}. We believe that whether such order may generally be single- or multi-\textbf{q} (which in turn depends on the lattice symmetry) warrants further investigation, because the electron itinerancy behind QMO formation also promotes higher-order spin interactions in an ionic picture. As discussed in the next section, such higher-order interactions may be intimately linked to the stabilization of multi-\textbf{q} order.

\section{3. Multi-\textbf{q} Order: Previous Examples}
In crystals with sufficiently high rotational symmetry, conventional single-\textbf{q} magnetic order can break the original rotational symmetry when the symmetry operations do not leave the ordering wave vector invariant. In these cases, a natural way to restore the symmetry is to conceptualize a multi-\textbf{q} variant of the magnetic order, which inherently respects the full rotational symmetry of the material's paramagnetic state. Multi-\textbf{q} magnetic order can be viewed as the vector sum of all symmetry-equivalent single-\textbf{q} components generated by the action of the material's rotational symmetry operations. This summation typically results in a spin configuration that is non-collinear or even non-coplanar, even when the spins in each individual single-\textbf{q} component are collinear (as long as they are not parallel to the rotational symmetry axis broken by the single-\textbf{q} order). This distinguishing feature sets multi-\textbf{q} magnetic order apart from conventional single-\textbf{q} structures, offering richer, often more complex, magnetic textures. Before we discuss the subject of multi-\textbf{q} magnetic order in the honeycomb cobaltates, it is useful to survey the formation mechanism and experimental determination of multi-\textbf{q} order in previously studied materials.

In magnetic Hamiltonian with bilinear spin interactions, single- and multi-\textbf{q} magnetic orders are often found to be degenerate in their semiclassical energy \cite{DiopPRB2022,ChenPRL2023}. As a result, thermal, quantum and quenched disorders \cite{HenleyPRL1989, ShengJOP1992, SchickPRB2020, DiopPRB2022, Liu2016PRB,Seabra2016PRB} become important factors for selecting the magnetic ground state. A well-known form of multi-\textbf{q} magnetic order is the skyrmion lattice \cite{Muhlbauer2009Science, Bogdanov1994JOM}, which has been found in a variety of cubic and hexagonal magnetic metals with potential for spintronics applications \cite{Kanazawa2011PRL, Nakatsuji2015Nature,TakagiSA2018}. Skyrmion lattices are believed to be stabilized by the joint effects of thermal fluctuations, external magnetic fields, Dzyaloshinskii-Moriya interactions, and magnetization amplitude variations in non-centrosymmetric metallic magnets \cite{Roessler2006Nature}.

Multi-\textbf{q} magnetic order has also been experimentally observed and theoretically studied in a variety of centrosymmetric systems. Representative materials include cubic \cite{IshiwataPRB2020,Paddison2024NPJ} and tetragonal systems \cite{AllredNatPhys2016,AllredPRB2015,TaddeiPRB2016,KawarazakiPRB2000,Khanh2022Zoology,Wood2023PRB,SaitoPRB2023} (notably, some of the multi-\textbf{q} order in tetragonal systems are collinear, with spins parallel to the $C_4$ axis). In these systems, due to the presence of inversion symmetry, Dzyaloshinskii-Moriya interactions are no longer a key driving force, and magnetic frustration and/or spin anisotropy may play an important role \cite{Okubo2011PRB,Okubo2012PRL, KamiyaPRX2014,HayamiPRB2016,Seabra2016PRB,HayamiPRB2019,HayamiPRB2021,DiopPRB2022, YambePRB2023}. In addition, the effectiveness of long-range (such as dipole-dipole or conduction electron-mediated) and higher-order spin interactions (such as biquadratic and multi-spin interactions) in stabilizing multi-\textbf{q} magnetic order becomes highlighted in recent studies \cite{YuArxiv2017,HayamiPRB2017, TakagiSA2018, FangJPCM2019, PaulNatComm2020, KatoPRB2022, Paddison2024NPJ, HayamiPRB2023, HayamiReview2024, PohlePRB2023, ChenPRL2023}. While higher-order interactions in metallic magnets can arise from the interplay between conduction electrons and local moments within a Kondo-lattice picture \cite{Martin2010PRL, Kato2010PRL, HayamiPRB2017,OzawaPRL2017}, they can also be understood as a natural consequence of higher-order expansion of the Hubbard model \cite{YangNJP2012,SwainPRB2016}. It therefore seems that the presence of higher-order interactions may be considered a universal driving force for the formation of multi-\textbf{q} magnetic order in both metallic and insulating systems.

\begin{figure}[!hb]
\includegraphics[width=3.2in]{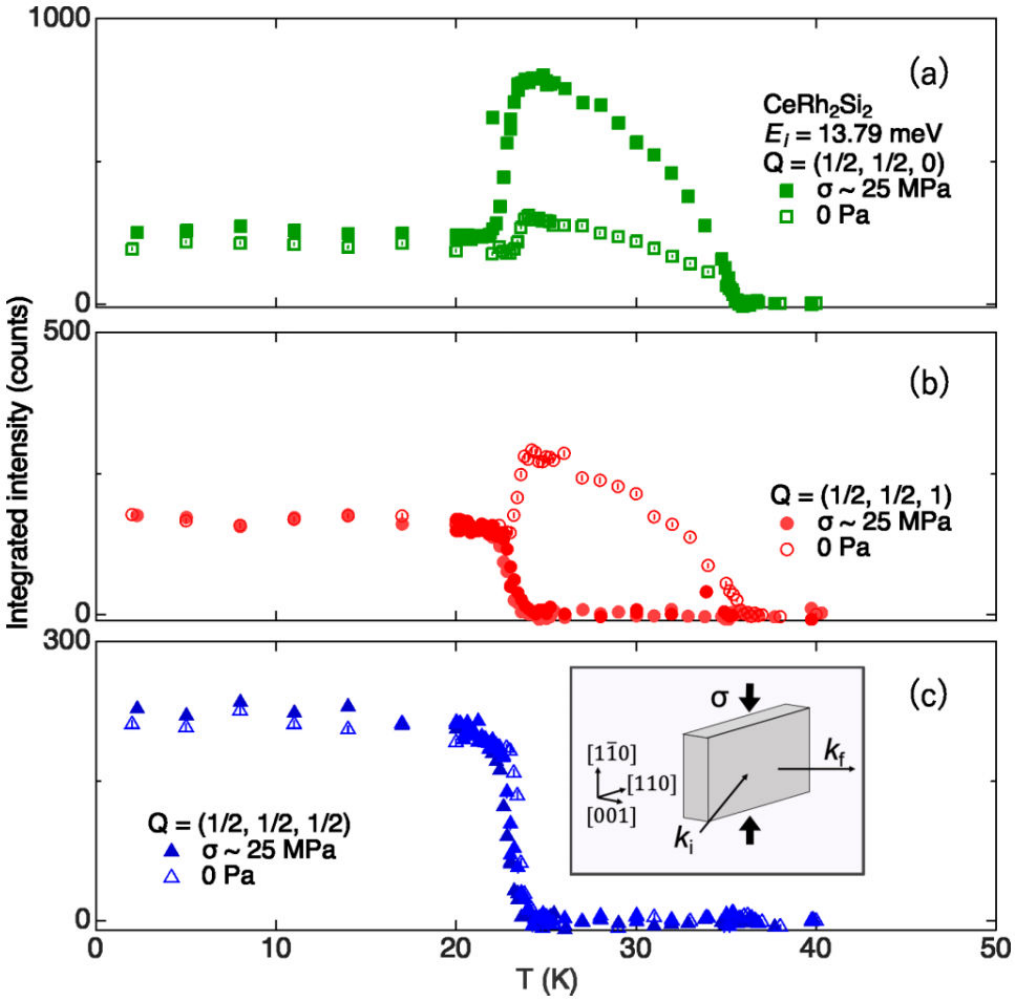}
\caption{Temperature dependence of magnetic neutron diffraction intensity in a single crystal of \ch{CeRh_2Si_2} (inset) cooled down with and without uniaxial stress, measured at various wave vectors: (a) $\mathbf{Q} = \mathbf{q} = (1/2, 1/2, 0)$, (b) $\mathbf{Q} = (1/2, 1/2, 1)$, which corresponds to $\mathbf{q} = (1/2, -1/2, 0)$ (measured from $\mathbf{G} = (0, 1, 1)$, due to the body-centered structure), and (c) $\mathbf{Q} = \mathbf{q} = (1/2, 1/2, 1/2)$. The first two $\mathbf{q}$'s have non-zero intensity in both the single- (between $T_\mathrm{N1}$ = 36 K and $T_\mathrm{N2}$ = 25 K) and multi-\textbf{q} magnetic states (below $T_\mathrm{N2}$), whereas the third $\mathbf{q}$ exclusively characterizes the multi-\textbf{q} state. The uniaxial compression $\sigma$, applied along the $(1, -1, 0)$ direction, favors the formation of the single-\textbf{q} domain in (a) at the cost of the one in (b), but does not strongly affect the multi-\textbf{q} order.  The figure is reproduced from \cite{SaitoPRB2023}.}
\label{fig2}
\end{figure}

Distinguishing between rotational symmetry-breaking single-\textbf{q} order and its symmetry-preserving multi-\textbf{q} variant can be challenging in conventional diffraction experiments \cite{KawarazakiPRL1988,ForganJOP1990}. This is because, in both scenarios, magnetic diffraction intensities at the rotational symmetry-related \textbf{q} vectors are expected to be nearly equal. The equality is intrinsic to multi-\textbf{q} order, but extrinsic in the case of single-\textbf{q} order due to randomized domain population. A viable strategy to differentiate the two is hence to create an unequal domain population in the latter case, by ``training'' the sample with external symmetry-breaking conditions. An example of this is displayed in Fig.~\ref{fig2}, where the magnetic diffraction peaks from a single crystal of \ch{CeRh_2Si_2} are compared after cooling the sample with and without uniaxial compression \cite{SaitoPRB2023}. The compression perturbs the symmetry of the Hamiltonian and cause single-\textbf{q} domains to have different likelihoods of formation, resulting in variations in the corresponding magnetic diffraction intensities [Fig.~\ref{fig2}(a-b)]. In contrast, the diffraction intensities associated with the symmetry-preserving multi-\textbf{q} order are relatively insensitive to the compression [Fig.~\ref{fig2}(c)]. In this way, the single- (between 25 K and 36 K) and multi-\textbf{q} (below 25 K) nature of the ordered states can be distinguished. Similar experiments using strain or magnetic fields to ``detwin'' single-\textbf{q} domains can be found for elucidating the multi-\textbf{q} nature of magnetic order in a variety of materials, \textit{i.e.}, based on the lack of detwinning effects \cite{NormilePRB2002,TakagiSA2018,Khanh2022Zoology,SaitoPRB2023}.

\begin{figure}[!ht]
\includegraphics[width=3.2in]{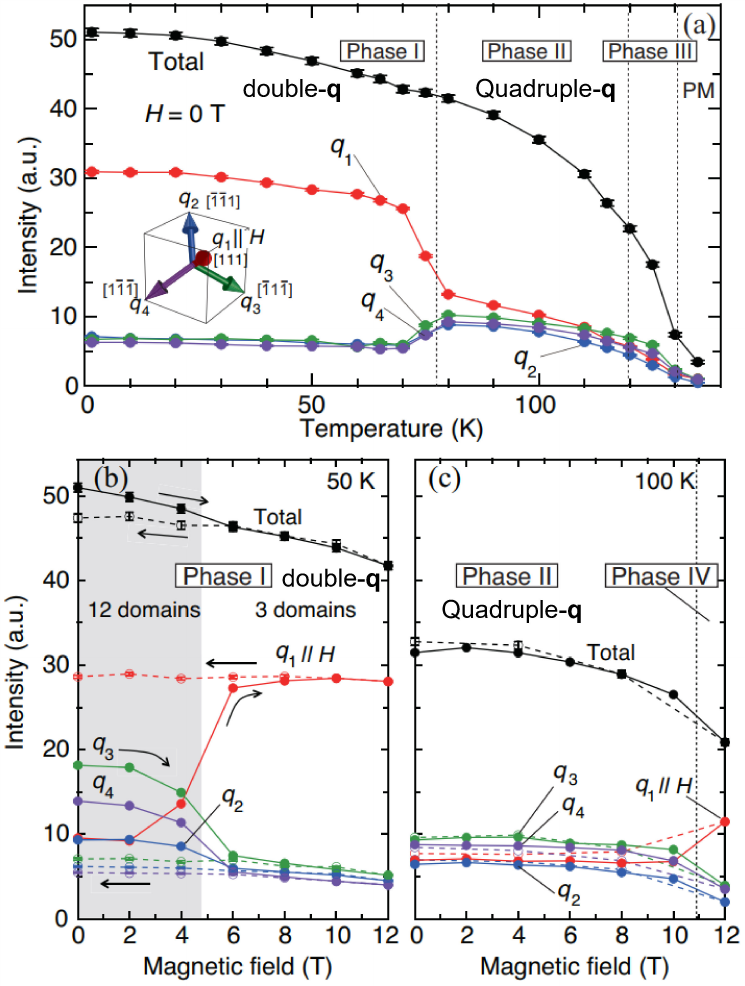}
\caption{(a) Temperature dependence of magnetic intensities measured at $\mathbf{q}_i$ ($i = 1$-4) in zero field, after field-cooling the sample to the lowest temperature. (b) and (c) Magnetic field dependence of the scattering intensities measured in the double-\textbf{q} phase I and the quadruple-\textbf{q} phase II, respectively. The data with solid lines and broken lines were measured on increasing and decreasing the field along [111], respectively. All the data were measured after zero-field cooling from room temperature. The figure is reproduced from \cite{IshiwataPRB2020}.}
\label{fig3}
\end{figure}

It is important to note that even a multi-\textbf{q} order may break some (but not all) of the rotational symmetry \cite{ForganJOP1990}. A notable example of this is found in cubic \ch{SrFeO_3} \cite{IshiwataPRB2020}, as demonstrated by the measurement data in Fig.~\ref{fig3}. Similar to the previous example, a magnetic field is applied upon cooling the sample, in order to alter the domain population indicated by the intensities of different magnetic Bragg peaks. The system is in a quadruple-\textbf{q} ordered state in phase II [Fig.~\ref{fig3}(a,c)], as manifested by the absence of clear field-training effects because the more stable quadruple-\textbf{q} order preserves the full rotational symmetry of the cubic crystal. In phase I below 80 K, however, the more stable magnetic order becomes double-\textbf{q}, which allows for the development of orientational domains that are trainable by the field. The training field can be applied either upon the initial cooling [Fig.~\ref{fig3}(a)], or at low temperatures after zero-field cooling [Fig.~\ref{fig3}(b)] as long as the field is sufficiently large, and the training effect persists even after the field is later removed. The training mechanism can be understood as follows. Without an external field, the double-\textbf{q} order naturally develops in a total of 12 randomly distributed domains, each featuring the combinations of two wave vectors among $q_1$, $q_2$, $q_3$ and $q_4$ (and with time-reversal twins). In a training field, one of the wave vectors (say $q_1$) becomes favored over the other three, such that three of the twelve domains ($q_1 + q_2$, $q_1 + q_3$, and $q_1+q_4$) are preferred. This makes the double-\textbf{q} order's diffraction intensities at the unfavored wave vectors remain non-zero and amount to about one third of that at the favored wave vector \cite{IshiwataPRB2020}.

Multi-\textbf{q} states can also be identified through specific characteristics accessible in diffraction measurements without training the sample. One approach is to search for higher-order harmonic peaks, which can be interpreted as the coherent superposition of distinct \textbf{q} components \cite{ForganPRL1989, Wood2023PRB}. However, this method may sometimes be confused with multiple scattering phenomena. Additionally, it has been suggested that under certain conditions, specific multi-\textbf{q} states may exhibit zero intensity at the positions of higher-order harmonic peaks \cite{HayamiPRB2023}. Another approach involves detecting associated structural symmetry breaking, which reflects magnetoelastic coupling effects in the rotational symmetry-breaking single-\textbf{q} state \cite{AllredPRB2015, TaddeiPRB2016}. Although high-resolution X-ray or neutron scattering techniques are required for this purpose due to the typically small induced lattice distortions, symmetry sensitive probes such as Raman scattering (for observing phonon splitting  \cite{Yue2017PRB}) may be used as complementary methods.

\begin{table*}[!t]
\begin{ruledtabular}
\renewcommand{\arraystretch}{1.3}
\begin{tabular}{lccccl}
- & Crystal structure & Type & Stabilized by & Way to identify & Refs. \\
\hline
\ch{Nd} &  hexagonal  & metal & N.A. & higher-order satellites & \cite{ForganPRL1989}\\
\ch{Mn-Ni} alloy & face-centered cubic  & metal & N.A. & $\gamma$-ray emission & \cite{KawarazakiPRL1988}\\
\ch{USb} & face-centered cubic  & metal & N.A. & excitation spectrum & \cite{Jensen1981PRB}\\
\ch{(U_{1-x}Pu_x)Sb} & face-centered cubic  & metal & N.A. & field detwinning & \cite{NormilePRB2002}\\
\ch{SrFeO_3} & simple cubic  & metal & thermal fluctuations & field detwinning & \cite{IshiwataPRB2020}\\
\ch{Y_3Co_8Sn_4} & honeycomb  & metal & four-spin interaction & field detwinning & \cite{TakagiSA2018}\\
\ch{MnTe2} & simple cubic  & semicoductor & DM interaction & M\"{o}ssbauer & \cite{Burlet1997PRB}\\
\ch{Co_{1/3}TaS_2} & triangular & metal & four-spin interaction & excitation spectrum & \cite{ParkNC2023, Park2024Arxiv}\\
\ch{Co_{1/3}NbS_2} & triangular & metal & four-spin interaction & M\"{o}ssbauer & \cite{Dong2024PRB}\\
\ch{CeRh_2Si_2} & tetragonal & metal & biquadratic interaction & stress detwinning & \cite{SaitoPRB2023}\\
\ch{GdRu_2Si_2} & tetragonal & metal & higher-order interaction & field detwinning  & \cite{Khanh2022Zoology}\\
& & & & higher-order satellites & \cite{Wood2023PRB} \\
\ch{(Sr_{1-x}Na_x/Ba_{1-x}K_x)Fe_2As_2} & tetragonal & superconductor & biquadratic interaction & diffraction \& M\"{o}ssbauer & \cite{AllredNatPhys2016,AllredPRB2015,TaddeiPRB2016}\\
\ch{Ba_2(Y/Lu)RuO_6} & cubic & insulator & biquadratic interaction & excitation spectrum & \cite{Paddison2024NPJ,FangJPCM2019}\\
\ch{Na_2Co_2TeO_6} & honeycomb & insulator & ring exchange & excitation spectrum & \cite{ChenPRB2021}\\
& & & & field detwinning & \cite{YaoPRR2023} \\
\ch{Na_3Co_2SbO_6} & monoclinic honeycomb & insulator & N.A. & field detwinning & \cite{Gu2024PRB}\\
\end{tabular}
\end{ruledtabular}
\caption{List of materials with reported multi-\textbf{q} magnetic states, along with the key experimental indications.}
\label{tab2}
\end{table*}

Further signatures of multi-\textbf{q} magnetic states may be found in the local spin orientations. As the electronic magnetic moments can affect the nuclear moments via the hyperfine coupling, they can be detected by nuclear technologies such as $\gamma$-ray emission \cite{KawarazakiPRL1988} and M{\"o}ssbauer spectroscopy \cite{Dong2024PRB, Burlet1997PRB, AllredNatPhys2016}. Moreover, in some special cases, multi-\textbf{q} states may be recognized through spin-polarized scanning tunnelling microscopy \cite{Gastiasoro2017NC,Spethmann2020PRL} and nuclear magnetic resonance (NMR) \cite{Tokunaga2005PRL}. In addition to static properties, multi-\textbf{q} states can also leave signatures in the spin dynamics, which can be measured by inelastic neutron scattering (INS) \cite{Jensen1981PRB, ParkNC2023, Paddison2024NPJ, Park2024Arxiv}. Evidence for multi-\textbf{q} states may be obtained by either fitting the excitation spectrum \cite{Jensen1981PRB, ParkNC2023, Paddison2024NPJ} or scrutinizing their characteristic behavior that are distinct between single- and multi-\textbf{q} states \cite{Park2024Arxiv}. Table~\ref{tab2} summarizes how multi-\textbf{q} magnetic ground states have been experimentally indicated in a variety of compounds.

\section{4. Evidence for multi-\textbf{q} order in honeycomb cobalt oxides}

As discussed in Section 2 and illustrated in Fig.~\ref{fig1}, the crystallographic symmetries of \ch{Na_2Co_2TeO_6} and \ch{Na_3Co_2SbO_6} are compatible with $M$-point magnetic order in either single- or multi-\textbf{q} form. To tell which of these scenarios is more likely to be correct, single-crystal neutron diffraction experiments have recently been performed with in-plane magnetic fields in both \ch{Na_2Co_2TeO_6} \cite{YaoPRR2023} and \ch{Na_3Co_2SbO_6} \cite{Gu2024PRB}. Here we only focus on the  training effects of the fields on the magnetic domains, as have been discussed in the previous Section.

\begin{figure*}[!t]
\includegraphics[width=6.9in]{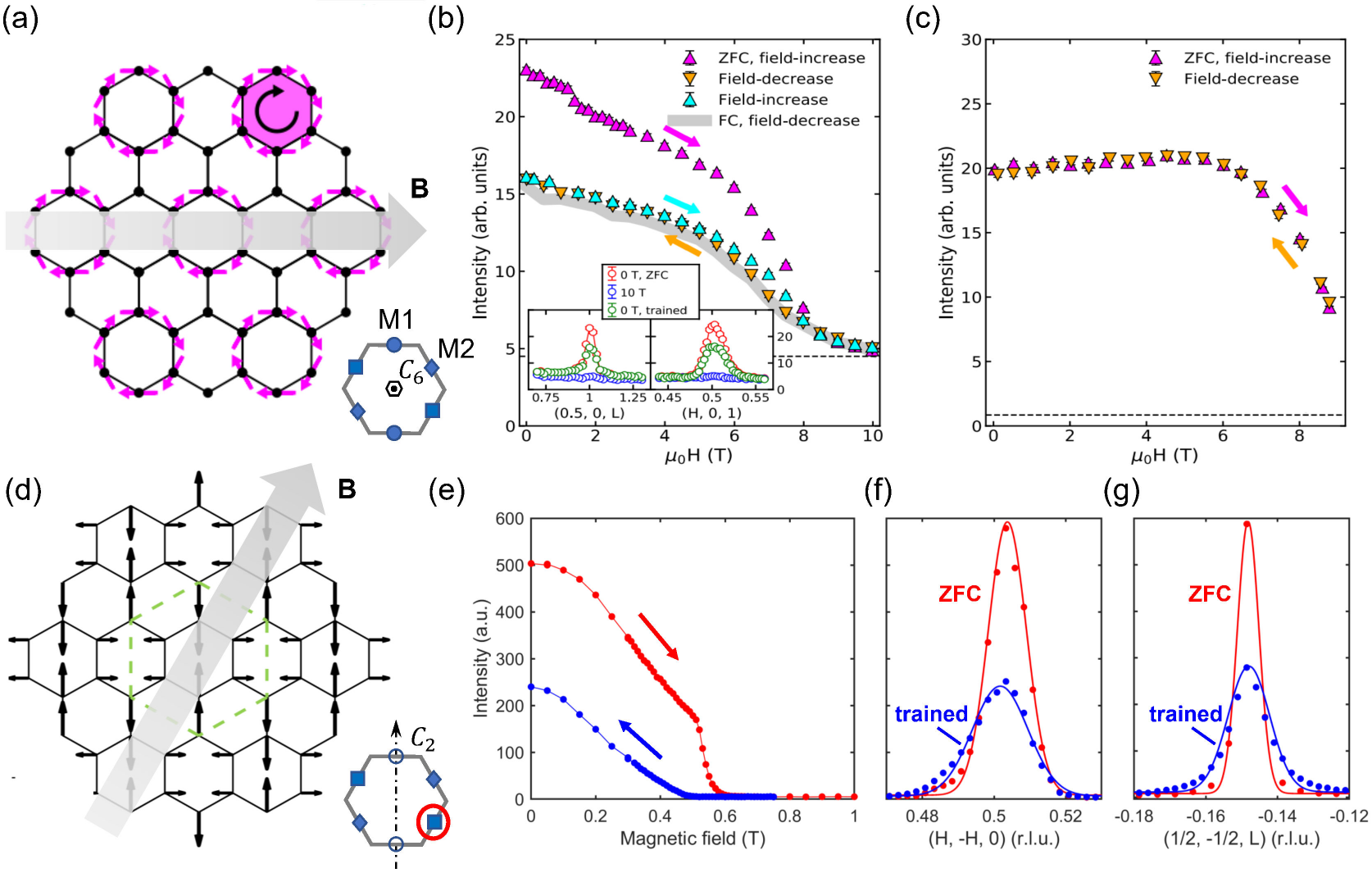}
\caption{(a) Schematic of triple-\textbf{q} magnetic structure in \ch{Na_2Co_2TeO_6}, constructed by summing up three zigzag components whose propagating wave vectors are different $M$-points of the Brillouin zone (inset). The direction of the applied magnetic field is unfavorable for the $M_1$ component. (b) and (c) Field evolution of magnetic Bragg peaks at $M_1 = (0.5, 0, 1)$ and $M_2 =(0, 0.5, 1)$ in \ch{Na_2Co_2TeO_6} at 2 K. Data displayed in the main panels are measured at the maximum of the peak profile (insets). Horizontal dashed lines indicate background level. (d) Schematic of double-\textbf{q} magnetic structure in \ch{Na_3Co_2SbO_6}, constructed by summing up two zigzag components that are related to each other by the $C_2$ rotational symmetry about the \textbf{b} axis. With a magnetic field applied along the indicated low-symmetry direction, the zigzag component (circled in the inset) measured in (e-g) is expected to be unfavored. (e-g) Field evolution and training effect of the magnetic Bragg peak at $(1/2, -1/2, 0)$ in \ch{Na_3Co_2SbO_6} at 2 K. The figures are adapted from \cite{YaoPRR2023} and \cite{Gu2024PRB}.}
\label{fig4}
\end{figure*}

In \ch{Na_2Co_2TeO_6}, the single-\textbf{q} magnetic structure (or single-\textbf{q} component in a multi-\textbf{q} structure) has been constrained \cite{LefrancoisPRB2016,BeraPRB2017} to have the in-plane component of magnetic moments in each zigzag chain ferromagnetically aligned parallel to the chain, \textit{i.e.}, perpendicular to \textbf{q}. The moments are further canted towards the \textbf{c} axis in an alternating fashion between the chains. Assuming that the multi-\textbf{q} variant of the magnetic structure preserves all of the crystallographic rotational symmetries, one can infer its in-plane magnetic moment arrangement by summing up three single-\textbf{q} components 120$^\circ$ from each other \cite{ChenPRB2021,YaoPRR2023}, which results in an ordering pattern with spin ``vortices'' as displayed in Fig.~\ref{fig4}(a).

In the recent experiments reported in Ref.~\onlinecite{YaoPRR2023}, the magnetic domains in \ch{Na_2Co_2TeO_6} were trained either at the lowest temperature (2 K) up to a maximal field of 10~T, or upon cooling from the paramagnetic state in a 10~T field, in a field geometry as shown in Fig.~\ref{fig4}(a). Figures~\ref{fig4}(b) and (c) present the evolution of magnetic Bragg peak intensities at $M_1 = (0.5, 0, 1)$ and $M_2 = (0, 0.5, 1)$, respectively, demonstrating the effect of training. As the training field was removed, the $M_1$ intensity recovered to about 2/3 of the value originally observed after zero-field cooling, whereas the $M_2$ intensity returned to the original value without a significant difference. These results showed that the training process had failed to eliminate the formation of any presumed, energetically unfavored single-\textbf{q} domains. While the 1/3 intensity decrease at $M_1$ might indicate a partial suppression of the corresponding single-\textbf{q} domain, it was found that the lost intensity (measured at integer $L=1$) was redistributed into rod-like magnetic diffuse scattering along the $\mathbf{c^*}$ axis, rather than into diffraction signals at other $M$ points. In other words, while the training was able to affect the magnetic correlation length across the honeycomb layers, there was no indication for a repopulation of single-\textbf{q} orientational domains within the layers \cite{YaoPRR2023}. These observations were in favor of a triple-\textbf{q} magnetic ground state in \ch{Na_2Co_2TeO_6}. The only remaining alternative possibility was that there existed unidentified structural pinning effects, such as frozen-in uniaxial strains, which are randomly distributed in the samples yet sufficiently strong to overrule the effects of the magnetic fields. This issue was addressed in a more recent study \cite{JinArxiv2024}, where an emergent effect arising from the spin vorticity [Fig.~\ref{fig4}(a)] was further observed with optical Faraday rotations.

\begin{figure*}[!t]
\includegraphics[width=5.5in]{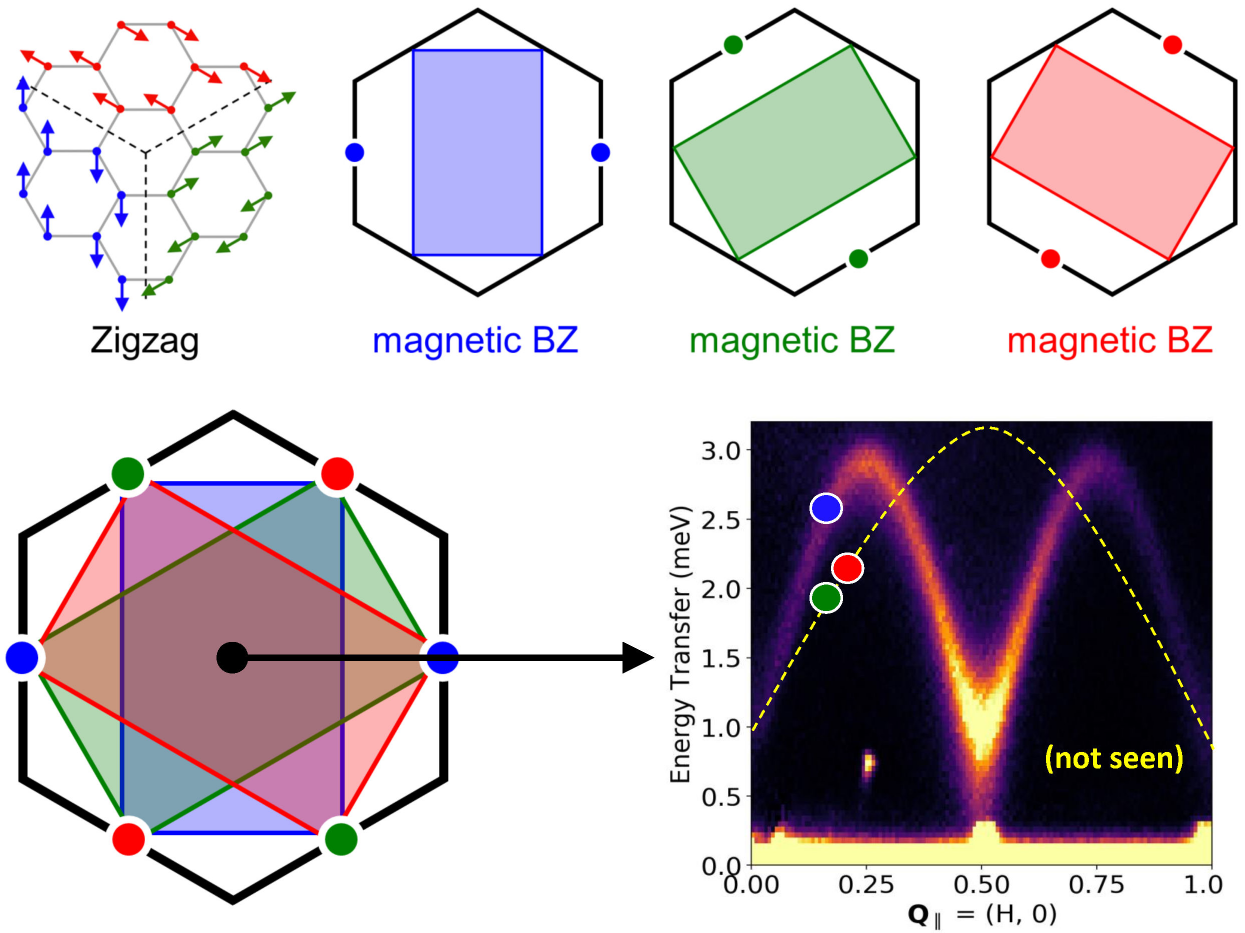}
\caption{Upper half: Three orientational domains of the zigzag order, along with their corresponding propagating wave vectors and magnetic Brillouin zones. Lower-left: Superposition of three types of magnetic Brillouin zones. When spin waves are measured along the momentum trajectory that connects two neighboring structural Brillouin zones (black arrow), two types of dispersion relations are expected to be seen, because the trajectory goes through magnetic zone centers three times in the ``blue'' domain, but only twice (plus zone corner once) in the ``green'' and ``red'' domains. Lower-right: Experimental spectra are consistent with the understanding that the momentum trajectory goes through magnetic zone centers three times, \textit{i.e.}, without seeing the presumed contributions from the ``green'' and ``red'' zigzag domains, hence contradicting the zigzag scenario but consistent with the triple-\textbf{q} scenario. The figure is adapted and modified from \cite{ChenPRB2021}.}
\label{fig5}
\end{figure*}

\begin{figure*}[!t]
\includegraphics[width=6.8in]{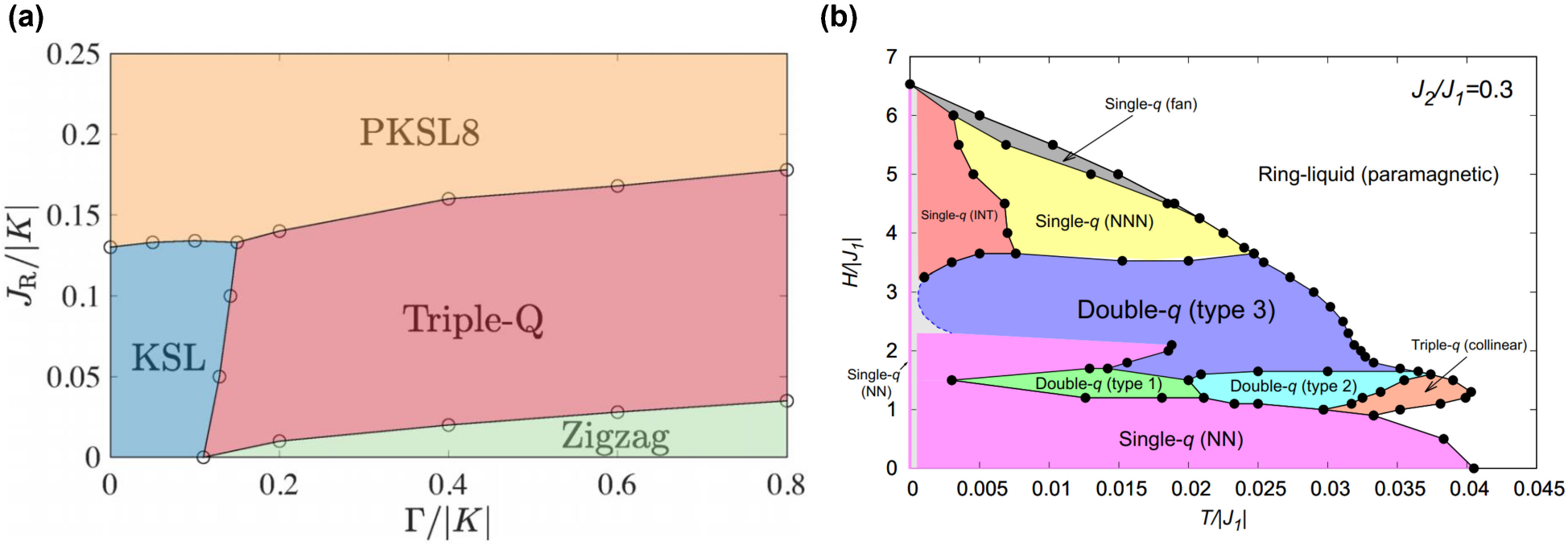}
\caption{(a) Phase diagram of the quantum $K-\Gamma-\Gamma^\prime-J_R$ model for $K$ \textless~0, $\Gamma$ \textgreater~0, $J_R$ \textgreater~0, and $\Gamma^\prime$/$\left|K\right|$ = -0.05 in the limit of large system size, calculated by variational Monte Carlo method. There are two different types of QSL phases: the KSL and the PKSL8, near the triple-\textbf{q} phase. (b) The $H-T$ phase diagram of the $J_1-J_2$ honeycomb-lattice Heisenberg model with $J_2/J_1$ = 0.3 determined by Monte Carlo simulation. The low-temperature light-gray region is the region where the thermalization
cannot be achieved, while the zero-temperature limit can be identified based on the low-temperature expansion as a single-\textbf{q} (NN) state. The dotted blue line representing the low-temperature phase boundary of the double-\textbf{q} (type 3) state remains somewhat arbitrary. This figure is adapted from \cite{Wang2023PRB} and \cite{Shimokawa2019PRB}.}
\label{fig6}
\end{figure*}

The experimental situation in \ch{Na_3Co_2SbO_6} is similar. Here, the question is between single- and double-\textbf{q} scenarios. The double-\textbf{q} magnetic ground state \cite{Gu2024PRB} features a non-collinear arrangement of the in-plane magnetic moments [Fig.~\ref{fig4}(d)], but it has no spin vorticity because of the absence of a third magnetic propagating vector along $\mathbf{b}^*$. To differentiate between the two scenarios, the in-plane training field was applied in a low-symmetry direction as illustrated in Fig.~\ref{fig4}(d). The measured magnetic Bragg peak [Fig.~\ref{fig4}(e-g)] was at the propagating wave vector which would correspond to domains that are unfavored by the training field in the single-\textbf{q} order scenario. The central finding was that the diffraction signal recovered after the training field was removed, even though the field was strong enough to fully suppress the order. Hence, the recovered signal was more likely a component of the full, double-\textbf{q} order parameter, rather than being associated with a given type of single-\textbf{q} domains. Similar to the case of \ch{Na_2Co_2TeO_6}, the peak intensity was observed to decrease after the training, due to a broadening of the momentum profile of the diffraction signals [Fig.~\ref{fig4}(f-g)] especially along the $\mathbf{c}^*$ axis. After accounting for the broadening, the total intensity was found to be unaffected by the training \cite{Gu2024PRB}.

A distinct characteristic of the multi-\textbf{q} ordering patterns in Figs.~\ref{fig4}(a) and (d) is that some of the sites are found to have smaller magnitudes of ordered moments than the rest. In  \ch{Na_2Co_2TeO_6}, the sites which have zero in-plane moments are expected to have ordered moments along the \textbf{c}-axis, which give rise to the system's ferrimagnetic behavior \cite{YaoPRB2020} in the ordered state, and which can be understood as related to the spin canting \cite{LefrancoisPRB2016,BeraPRB2017} in the individual zigzag components. However, the magnitude of these moments are not expected to be equal to the in-plane ones, because the system has pronounced overall easy-plane response to external fields \cite{YaoPRB2020}. In \ch{Na_3Co_2SbO_6}, the zero-field magnetic structure has been fully determined through spin-polarized neutron diffraction \cite{Gu2024PRB}, with varying moments on the lattice sites as shown in Fig.~\ref{fig4}(d). The presence of these reduced ordered moments implies the persistence of spin dynamics deeply in the ordered state due to the systems' magnetic frustration and quantum fluctuations, which has been confirmed in recent muon spin relaxation measurements \cite{Miao2024PRB}.

The above evidence for multi-\textbf{q} magnetic order in \ch{Na_2Co_2TeO_6} and  \ch{Na_3Co_2SbO_6} came from diffraction experiments. Historically, the likelihood of triple-\textbf{q} order in \ch{Na_2Co_2TeO_6} was first revealed by inelastic neutron scattering experiments based on symmetry arguments \cite{ChenPRB2021}. As illustrated in the upper half of Fig.~\ref{fig5}, in the zigzag scenario, the three orientational domains feature rectangular magnetic Brillouin zones that are rotated $120^\circ$ from each other. As a result, in a large, co-aligned sample for inelastic neutron scattering, one expects to observe spin excitations from all three types of domains simultaneously, \textit{e.g.}, with two types of dispersion relations seen on the momentum trajectory displayed in the lower half of Fig.~\ref{fig5}. The experimental spectra \cite{ChenPRB2021} strongly contradict this expectation, as only one branch of spin-wave dispersion is observed below 3~meV. The data are instead fully consistent with a triple-\textbf{q} understanding of the magnetic order, which has a smaller hexagonal magnetic Brillouin zone (which corresponds to the overlapped colored region in the lower-left of Fig.~\ref{fig5}).

Yet, it is important to note that such symmetry arguments alone might not be rigorous, as they cannot be used to rule out zigzag order formation under certain interaction parameter settings. For instance, when the interaction model has only Heisenberg interactions between the third-nearest neighbors ($J_3$), in which case the zigzag and triple-$\mathbf{q}$ orders are degenerate classical ground states, the magnon dispersions of different zigzag domains are also degenerate. This degeneracy arises because the spin lattice is effectively decoupled into different sublattices, as detailed in a discussion presented in Ref.~\cite{YaoPRL2022}. Such degeneracy is, however, not symmetry-enforced, as the absence of interactions between nearer neighbors than $J_3$ is a special requirement. When such interactions are present, the fitting of entire INS spectrum \cite{YaoPRL2022} (not just the lowest-energy branch) starting from a zigzag ground state, and considering the coexistence of multiple domains, has remained a challenge \cite{SongvilayPRB2020, LinNC2021, KimJOP2021}. Spin-wave fitting based on a triple-$\mathbf{q}$ ground state has yielded considerably more satisfactory results \cite{ChenPRL2023}, further supporting the triple-\textbf{q} scenario.

In addition to the neutron scattering efforts discussed above, indirect evidence for multi-\textbf{q} magnetic order has been obtained by other experimental techniques. These include null results in Raman scattering detection of phonon energy splitting in \ch{Na_2Co_2TeO_6} \cite{ChenPRB2021}, which ought to be present if the magnetic order (in conjunction with magnetoelastic coupling) strongly breaks the lattice $C_6$ rotational symmetry about the \textbf{c} axis, as well as nuclear magnetic resonance detection of the local internal fields in both \ch{Na_2Co_2TeO_6} \cite{LeePRB2021} and \ch{Na_3Co_2SbO_6} \cite{Hu2024PRB}: a multi-\textbf{q} magnetic ground state is expected to produce more complicated nuclear magnetic resonance spectra than a single-\textbf{q} one.

\section{5. Implications for theoretical models}

Information about the magnetic ground state imposes strong constraints on microscopic magnetic model construction, providing critical insights into the underlying interactions. As discussed in Section.~3, the multi-\textbf{q} order in centrosymmetric itinerant magnets highlights the significant role of long-range and higher-order interactions, which in those cases naturally arise from the interplay between local moments and conduction electrons. In contrast, conventional models for Kitaev-candidate honeycomb cobaltates are based on the tight-binding Hubbard insulator framework, which in first-order approximation translates into spin Hamiltonians with only bilinear spin interactions (\textit{e.g.}, Heisenberg and Kitaev interactions between two spins) \cite{SinghPRL2012, LiuPRB2018, SanoPRB2018, LiuPRL2020}. Within these models, the semi-classical energies of single- and multi-\textbf{q} orders are usually degenerate \cite{DiopPRB2022, ChenPRL2023}, but as displayed in Fig.~\ref{fig6}, the ground state can become multi-\textbf{q} order when higher-order spin interactions, or magnetic fields, are introduced \cite{JanssenPRL2016, Chern2017PRB, Shimokawa2019PRB, Wang2023PRB}. Indeed, in the case of \ch{Na_2Co_2TeO_6}, a triple-\textbf{q} ground state has been proposed to be stable in the presence of six-spin ring-exchange interactions, when the system is near a hidden SU(2)-symmetric point in the parameter space of an extended Heisenberg-Kitaev-Gamma model \cite{ChenPRL2023}. Notably, the model is also able to provide a satisfactory description of the full experimental spin-wave spectrum.

Such multi-spin, higher-order interactions are considered to emerge from higher-order expansions of the tight-binding Hubbard model \cite{ChenPRL2023}, and their presence underscores the critical role of electron hopping in Mott insulators. The significance of hopping be indicated by the dimensionless quantity $t/U$, where $t$ and $U$ are the inter-site hopping and on-site repulsion energies, respectively. Traditionally, hopping is expected to be less significant and more short-ranged in the cobalt oxides than in extended 5$d$- and 4$d$-electron systems \cite{LiuPRB2018, SanoPRB2018, LiuPRL2020}. However, recent detailed analysis of the electronic structure in one of the candidate materials, \ch{BaCo_2(AsO_4)_2} \cite{WinterJOP2022}, suggests that the hopping energy may have been underestimated. Specifically, the direct $xy/xy$ hopping $t' \sim$ -300 meV dominates over superexchange hoppings between $xz$ and $yz$ orbitals via ligand $p$ states $t \sim$ 50 meV. Similar findings were also reported for \ch{Na_2BaCo(PO_4)_2} and other edge-sharing Co oxides \cite{Wellm2021PRB, WinterJOP2022, Nair2018PRB}. This contradicts previous assumptions that ligand $p$ state-mediated hopping is predominant in the cobaltates \cite{LiuPRL2020} as a key factor behind dominant Kitaev interactions. Another significant deviation is the notable hopping between the third nearest neighbors, which becomes substantial already at the DFT level \cite{Maksimov2022PRB}. According to inelastic neutron scattering results, the presence of third-nearest-neighbour spin interaction also appears to be a general characteristic of materials such as \ch{BaCo_2(AsO_4)_2}, \ch{Na_2Co_2TeO_6} and \ch{Na_3Co_2SbO_6} \cite{HalloranPNAS2023, YaoPRL2022, ChenPRL2023, SongvilayPRB2020, KimJOP2021}.

The validity of microscopic models and their ground states can be further examined via the models' ability to explain related experimental observations. It has been reported for \ch{Na_2Co_2TeO_6} that, before developing the well-known 3D magnetic order, the system exhibits an intermediate phase featuring an unusual form of 2D order \cite{ChenPRB2021}. This phase has recently been explained using a model with ring-exchange interactions \cite{Francini2024PRB}. Depending on the sign of the ring-exchange interactions, either a zigzag or a triple-\textbf{q} ground state is favored, and they have intermediate ``vestigial'' phases in the forms of $\mathbb{Z}_3$ spin-nematic or $\mathbb{Z}_4$ spin-current density wave order, respectively. The experimentally observed vestigial phase in \ch{Na_2Co_2TeO_6} is consistent with the $\mathbb{Z}_4$ spin-current density wave order \cite{Francini2024PRB}. The order breaks lattice translational symmetry but preserves time-reversal symmetry, which explains why it does not lead to pronounced anomalies in magnetic susceptibility or NMR measurements \cite{YaoPRB2020, LeePRB2021, LefrancoisPRB2016, BeraPRB2017}, but can be observed in specific heat and diffraction measurements \cite{ChenPRB2021, YaoPRR2023}. The same model can explain the experimentally observed ferrimagnetic net moment in the 3D ordered phase as well \cite{YaoPRB2020,Francini2024PRB2}.

\section{6. Summary \& outlook}

To summarize, we have reviewed how the possible existence of multi-\textbf{q} order in magnetic materials can be experimentally addressed. This can be done either with reciprocal-space methods such as diffraction under external fields, with real-space methods such as NMR and STM. Current experimental evidence for such multi-\textbf{q} magnetic order in the honeycomb cobaltates has followed these routes. The findings extend the relevance of multi-\textbf{q} type of ordering from itinerant magnets to Mott insulators, promoting a re-examination of previously determined magnetic orders based on single-\textbf{q} ansatz in similar systems. Since the presence of higher-order and long-range interactions probably plays a crucial role in stabilizing the multi-\textbf{q} order, the results highlight the significance of electron itinerancy even in Mott insulator systems, challenging the conventional theoretical framework which leads to the prediction of dominant Kitaev interactions in the 3$d$ honeycomb cobalt oxides.

Despite their likely substantial deviations from the Kitaev honeycomb model, the cobalt oxides retain the potential to host exotic quantum states. As displayed in Fig.~\ref{fig6}(a), various spin liquid phases are predicted to emerge near the triple-\textbf{q} phase \cite{Wang2023PRB}. Moreover, in these nearly ideal two-dimensional honeycomb magnets, theoretical studies suggest that a spin liquid ground state might emerge from competing interactions, particularly between the first and third nearest neighbors \cite{HalloranPNAS2023, Merino2018PRB, Zhu2014PLB, Bishop2012JOP, Bose2023PRB}. The observation of a continuum scattering signal in \ch{BaCo_2(AsO_4)_2} via THz spectroscopy under a large \textbf{c}-axis magnetic field may be indicative of such a state \cite{Shi2021PRB, Zhang2023NM}. The inferred unequal-magnitude spin distribution in the multi-\textbf{q} states of \ch{Na_2Co_2TeO_6} and \ch{Na_3Co_2SbO_6} suggests that the classical order is suppressed by pronounced quantum fluctuations, a finding corroborated by muon spin relaxation measurements \cite{Miao2024PRB, Jiao2024CM}. While these materials finally develop antiferromagnetic order at low temperatures, the systems' tendency to form ferromagnetic correlations is not falling far behind and can be enhanced by external fields. Near the critical points between the quantum phases featuring these different magnetic correlations, exotic states with enhanced quantum fluctuations may emerge. In parallel to this optimistic view, it is equally important to acknowledge that seemingly novel observations might stem from relatively trivial origins such as disorders and lattice dynamics \cite{Lefrancois2022PRX, Lefrancois2023PRB, Sears2023PRB, Hong2024NPJ, Takeda2022PRR}. This highlights the need for careful interpretation of experimental data to distinguish genuine quantum phenomena from extrinsic effects.

\begin{acknowledgments}
We are grateful to our coworkers and colleagues for many fruitful collaboration and stimulating discu
ssions. The work at Peking University was supported by the National Basic Research Program of China (Grant No. 2021YFA1401901) and the NSF of China (Grant No. 12474138).
\end{acknowledgments}

\bibliography{cobaltate_review}

\end{document}